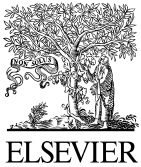

46th SME North American Manufacturing Research Conference, NAMRC 46, Texas, USA

# A Case Study for Blockchain in Manufacturing: "FabRec": A Prototype for Peer-to-Peer Network of Manufacturing Nodes

Atin Angrish[a], Benjamin Craver[a], Mahmud Hasan[a], Binil Starly[a,b,*]

*a Edward P. Fitts Department of Industrial and Systems Engineering*
*b Comparative Medicine Institute*
*North Carolina State University, Raleigh, NC USA 27695*

* Corresponding author. Tel.: +1 919 515 1815; fax: +1-919-515-5281 .
  E-mail address: bstarly@ncsu.edu

## Abstract

With product customization an emerging business opportunity, organizations must find ways to collaborate and enable sharing of information in an inherently trust-less network. In this paper, we propose – "FabRec": a decentralized approach to handle manufacturing information generated by various organizations using blockchain technology. We propose a system in which a decentralized network of manufacturing machines and computing nodes can enable automated transparency of an organization's capability, third party verification of such capability through a trail of past historic events and automated mechanisms to drive paperless contracts between participants using 'smart contracts'. Our system decentralizes critical information about the manufacturer and makes it available on a peer-to-peer network composed of fiduciary nodes to ensure transparency and data provenance through a verifiable audit trail. We present a testbed platform through a combination of manufacturing machines, system-on-chip platforms and computing nodes to demonstrate mechanisms through which a consortium of disparate organizations can communicate through a decentralized network. Our prototype testbed demonstrates the value of computer code residing on a decentralized network for verification of information on the blockchain and ways in which actions can be autonomously initiated in the physical world. This paper intends to expose system elements in preparation for much larger field tests through the working prototype and discusses the future potential of blockchain for manufacturing IT.

## 1. Introduction

In a Cybermanufacturing (CM) environment, the notion of networked organizations and machines creates an opportunity for product designs to be rapidly generated and then manufactured with limited human intervention through reuse of computer generated code or through automatic compilers that translate design data to machine instructions. However, ensuring trustworthiness among system elements in a Cybermanufacturing scenario can be challenging, particularly when there are multiple parties involved in the design, fabrication, production and verification of product assemblies. Traditionally this trust has been established between clients and manufacturers through extensive contract negotiation, acknowledgement of past historic performance, updated certifications and audits to ensure compliance. This 'trust tax' or rather the costs associated with ensuring trust among all parties in a supply chain of a product assembly is embedded into the final sales price charged to the customer [1]. In Cybermanufacturing, it is envisioned that we will have rapid commissioning and decommissioning of system elements, which means significant time and cost will need to be invested into securing trust in a networked environment spanning multiple business organizations. Traditional practices in establishing and evaluating trust will challenge the economic feasibility of cybermanufacturing platforms.

A partial solution to address this problem of trust determination is through the use of 'pull' based manufacturing platforms such as services provided by Li and Fung [2] or discrete part prototyping platforms such as those provided by Xometry [3], Maketime [4], Fictiv [5] and Plethora [6]. The digitalization and globalization has brought in an emerging era of pull-manufacturing eco-systems by connecting consumers with the capabilities of manufacturers and making them available to their customers. With near zero physical assets, these platforms are able to add significant value for platform participants, customers and service providers, by making information available to both parties involved. Consumers trust these platforms to bring in verified manufacturing service providers and lower the cost to sourcing parts from hundreds to thousands of service providers. However such centralized clearing-house type platforms will not make sense in highly regulated markets such as medical, aerospace and military manufacturing, even in the context of short-run production parts. For instance, a medical device startup company will want to know where the manufacturing of its products are carried out since it is an inherent aspect of obtaining FDA clearance on their product. In this scenario, these centralized platforms cannot hide information from each other as it is critical to know the identity and capabilities of service providers.

An alternate implementation would be to build platforms that are decentralized where each partner derives their fair share of value generated from participating in the platform ecosystem. The transparency offered by decentralization and the value offered through distribution of resource allocation can perhaps address the issue of trust determination in a cybermanufacturing network. Each organization inherently trusts only their own processes put in place. Communication across IT systems will require extensive software adaptors and agents to be written for interaction between organizations. This leads to redundant and outdated code that must be constantly rewritten and updated when rapid reconfiguration of system elements must take place.

A solution for enabling such a decentralization mechanism that is rapidly gaining attention from academia and industrial organizations is the concept of a shareable ledger that runs through a permissioned network, called the Blockchain mechanism [7]. This shareable ledger is not owned by any central authority and any authorized participant can view and write to its contents. Qualitative events related to a manufacturer are shared across a network of nodes and the data stored on the blockchain is immutable, hence establishing manufacturing data provenance. Such quantitative provenance was previously hard or expensive to obtain. Any manufacturer can make their organization's data accessible to any other participant on the network (ex. a client) to help establish an organization's reputation and hence partially determine trust between two parties. Several white papers have been released with adaptations to the data architecture now being explored in the finance, insurance, agriculture and electronic health records industry [8]. Basic implementations in manufacturing have also emerged in the context of decentralizing 3D printing resource availability and in reevaluating how data integration across manufacturing supply chains are implemented [9]. But this technology is still in its infancy and in our view, there are many manufacturing use case scenarios put forward with Blockchain technology that could otherwise be just as well implemented with traditional manufacturing information systems. Decentralized networks have been studied extensively in the past and have their own vulnerabilities. However, recent advancements in distributed computing, internet of things and advances in data analytics has provided a new impetus to reconsidering decentralized networks for manufacturing operations.

Current implementations of enterprise blockchain technology rely on the basic blockchain infrastructure provided by the Ethereum Foundation [10], EOS.IO [11], IBM HyperLedger and Microsoft CoCo frameworks [12], as well as improved blockchain database technology such as BigChainDB [13]. Much of the core infrastructure required to drive manufacturing applications are still in its infancy and will require significant development in terms of enabling machines and their associated systems to truly interact in a decentralized manufacturing ecosystem. New technology must be developed to authenticate the identity of machines on a network, new processes through which events recorded in the distributed ledger are verified and reconciled among the nodes, definition of the various roles and responsibilities of various nodes on the network, and studying mechanism designs that will power the interaction between the various levels of participants in the network of smart manufacturing marketplaces.

This paper focuses on a system framework through which a decentralized network of users and service providers can operate in a decentralized manufacturing eco-system, called 'FabRec'. We have highlighted specific use cases for which we think the Blockchain technology makes sense and why this would be a game-changer in terms of how manufacturing services are provided compared to current practices in sourcing service providers. We have implemented a portion of the infrastructure through a laboratory based setup of decentralized nodes comprising of computing nodes, a system-on-chip board and a physical CNC manufacturing machine. We demonstrate the setup of a smart contract among the nodes, to showcase how machines on a decentralized network can autonomously interact without human involvement. Through this relatively simple contract, we demonstrate how verification of the event is achieved and the mechanism of the consensus algorithm used to ensure blockchain integrity. This smart contract also demonstrates how instructions can be sent to the physical world once conditions are met within the code based contract. Lastly, we discuss the various technology pieces that must be developed from a manufacturing and computer science perspective to help move this nascent infrastructure technology to mainstream manufacturing.

## 2. System Implementation

### 2.1. FabRec System Overview

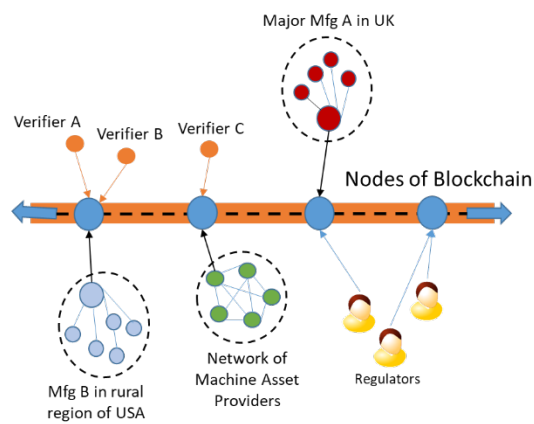

Figure 1: FabRec Overview. A sample snapshot of manufacturers, designers, asset builders and regulators participating on a decentralized network. This network enables controlled information sharing across trust boundaries (dotted circles).

In FabRec, each participant on the decentralized network will have various roles and responsibilities which allows them to interact with other members of the network. A participant can be a human, manufacturing machine, a computing node or an agent representing an organization, with each participant having an address identifying the participant on the network. FabRec, while still being decentralized will have a functional organization to the various participants on the network. Figure 1, highlights a sample view of a decentralized manufacturing eco-system for the purposes of acquiring manufacturing job services from across the open market. Each main node on the network has a copy of all blocks as part of the network. The dashed lines indicate boundaries of trust that are typical in current IT installations. Data contained in the blockchain allows integration of data across these trust boundaries.

Each participant in the network has a unique address which identifies the participant on the network (Figure 1). Manufacturers in any part of the world with verified capabilities gain access to the network. Access is also granted to significant machine assets currently owned by the manufacturer. Other participants include verifiers such as ISO body quality certifiers that can independently verify the validity of a certification claim made by a manufacturer. Networks of machine asset builders can be part of the network either contributing data or simply gaining access to the data made available on the blockchain network. Users, such as designers who are seeking job shop service providers can be participants accessing information such as verifying the capabilities of a particular manufacturer or placing request for service claim on the network. Similarly, regulators and compliance agents can also verify claims made by manufacturers particularly in relation to regulated products such as medical implants, aerospace components etc.

## 2.2. Blockchain Fundamentals

We cover the basic elements of the core infrastructure behind the Blockchain technology, particularly in relation with recent developments to building Blockchain 3.0 [14]. The brief description is provided to introduce terms used throughout the paper.

### 2.2.1. Block and Its Data Payload

Blockchain is a distributed software mechanism that provides a system with a continuously growing list of trusted asset transactions without the need for a central trust authority [15]. Such transactions are stored in a block-chain – which are essentially linked data structures, containing a batch of valid and verified transactions. It can be considered to be a continuously growing ledger that retains a permanent record of all the transactions that have taken place in chronological order and whose contents are immutable. Each block consists of an immutable hash of the prior block that it is connected to, which ultimately forms a chain link of blocks containing data that can be uniquely associated with a physical asset, such as a person or a physical property. This distributed database runs on multiple servers (nodes) across an entire network, with each node verifying the security and integrity of the data entry in blocks within this peer-to-peer network (Figure 1). Since there is no central system, trust is distributed across the nodes in the blockchain network.

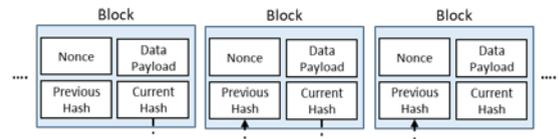

Figure 2: Linked data blocks with content hashed to detect modification. Chain of blocks leads to bread crumb trail of events linked to manufacturing transactions across participants in the network.

A block in a blockchain is a record of some or all the recent transactions that have taken place over the network. A transaction can contain recordings of significant and relevant events related to the asset being tracked. A block structure contains information about the most recent transactions, the size of the block, a transaction counter that keeps a track on the block instance being added to the blockchain, a block header retaining information on the cryptographic hash of the previous and the current block, the current timestamp and a 'number used only once' or a Nonce random number that aids in the generation of valid hashes for subsequent blocks as shown in Figure 2 [16-17].

### 2.2.2. Consensus Algorithms

Consensus algorithms are critical to the operation of decentralized and distributed systems, as it strives to achieve consistency of data among all the nodes on a decentralized system. The responsibility for the verification and analysis of data on a decentralized system can be undertaken by any arbitrary or assigned node on the network. To solve this problem in a trustless environment, it would make sense that every node of the network carries the exact same copy of the record of each transaction. This way, as long as the majority of users do not collude, the decentralized system will maintain a trustworthy environment for parties to work together without the need for a third-party facilitator. This process of enabling network participants to act as verifiers for transactions in exchange for rewards is called as mining. Blockchain mining is one of the most important paradigms in the idea of decentralized and distributed systems. Mining within the context of blockchain implies a mechanism for establishing consensus on the state of the blockchain to ensure that it is both secure and tamper-resistant. The process of mining is crucial to understanding the blockchain paradigm and immutability of its data contents. One of the most popular blockchain implementation, the bitcoin blockchain implements a mining algorithm similar to the hashcash algorithm, called the Proof of Work consensus algorithm [18].

Apart from Proof of Work (PoW), there are several other mining algorithms - Proof of Stake and Proof of Authority (PoA). Proof of stake [19] proposed by King and Nadal for Peercoin do not rely on traditional mining processes, rather it proposes selection of the verifiers based on random factors of parameters at "stake" such as wealth or reputation combined with hash values etc. Proof of authority (PoA) is proposed as a way of implementing consensus in a private network where the participants are vetted and aware of each other's identity [20]. The system is run by N trusted nodes (authorities) who have the authority to verify new transactions (or data) and issue new blocks to be added to the blockchain. A minimum number of authorities are assumed to be honest (= N/2 + 1 authorities). If there is a conflict or fraud, it is possible to stop a particular authority from participating in the new mining processes. Both PoW and PoA has been interchangeably used in our prototype implementation.

### 2.3. Block Architecture at the Manufacturer Level

A manufacturing service provider will have under its ownership multiple manufacturing machines providing the necessary capability and capacity to serve its customers. Each of these machines with a unique identity will be given the ability to directly write critical 'events' to the private blockchain in which the manufacturer participates in. Events recorded by the machine are transferred to its virtual twin [21], through which events are batched up and served up to the blockchain database. The content of the data block or the 'payload' of the block will depend on the following 1) Type of machine; and 2) Type of event being recorded, and 3) Number of transactions being recorded into the block. Hence one block can contain multiple transactions related to the operation of a machine.

Software adaptors are written to record significant events related to the machine on to the blockchain based distributed database. These events or manufacturing data transactions can include several types such as - 1) Machine Asset Information, such as installation, service, machine upgrades, break-down events, access log over the duration of its lifetime etc.; 2) Machine Utilization can include recording events such as overall equipment effectiveness (OEE), up-time, maintenance logs, power consumption; and 3) Machine Capability, such as materials worked with, complexity of features, raw workpiece type etc. An example data structure is presented in Figure 3, that comprise the Block with an event related to a Machine Asset type. As a particular machine is utilized by the manufacturing shop over time, these events are recorded into the chain and a breadcrumb trail created. This chain creation is repeated for all the machines under the ownership of the manufacturing service provider. This data is broadcast to all nodes on the blockchain network for verification and validation. Hence, when a client intends to do business with a job shop service provider, a permissioned access is granted to the client. The potential client's agent is able to then track historical performance data of the manufacturer and can verify the authenticity and provenance of the data. Such information sharing can increase the reputation of the service provider and improves the odds of winning new business.

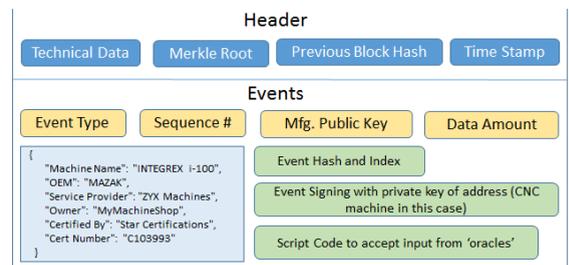

Figure 3: This example block displays the certification of a machine's capabilities by a neutral 3rd party. This 'Block' maintains a chain-link to all events generated by the specific machine in the past

and all future events that will be created during its lifetime

Two data models are developed that is central to the development of the decentralized manufacturing data blockchain, particular in the context of establishing trust between two parties. They are 1) Transaction Model; and 2) Block Model. The following section, describes our proposed data models for each model type.

1. The transaction model is the most basic kind of record stored within the distributed ledger system. We focus on 3 types of manufacturing data transactions: 1) Machine Asset; 2) Machine Utilization; and, 3) Manufacturing Capability. A machine asset transaction can include creation of a machine asset, upgrades to the asset, service updates or discontinuation of an asset. It can include description of the asset type, its capabilities, owners of the asset, its service providers, a list of human operators etc. A Machine utilization transaction can include description of overall equipment effectiveness (OEE), up-time, maintenance logs, scrap rates, power consumption and other meta-data with regards to the utilization of the machine asset over time; and 3) Machine Capability, such as materials worked with, complexity of features built and tolerances held etc. All of these transaction types may be represented by a JSON-like document with the following structure:

```
{
    "id": "<hash of transaction, excluding signatures to be the database primary key >",
    "version ": "<version number of transaction model to support future changes>",
    "transaction ": {
        " Service Provider ID ": " <public key of service provider >",
        " Machine ID ": "<public key of machine asset>",
        "operation": "< transaction type of either Create/Utilization/Capability>",
        " timestamp ": " < timestamp when transaction was submitted>",
        " data ": {
            " hash": " < the hash of the data payload described below >",
            " payload" : " <an embedded JSON document with a format dependent on operation type >
        }
    }
}
```

Figure 4: Transaction Model

2. The Block Model comprises a list of transactions that will be aggregated, verified, validated and signed by the federation of nodes. Its structure is as follows:

```
{
    "id": "<hash of the block and will be a primary key to ensure blocks are unique>",
    "prev_block ": " <hash of the previous block in the chain>",
    " block ": {
        " timestamp ": " < Block creation timestamp by the leader node >",
        " transactions ": ["<list of transactions included in the block>"],
        " node_pubkey ": " < the public key of the leader node that created the block >",
        " voters ": [" < list of public keys of nodes who voted to validate the block >"]
    },

    " signature ": " < signature of the block by the leader that created the block. The signature is created by
        serializing the contents and signing with its private key >",

    " votes ": [" < list of votes provided by other federation nodes > "]
}
```

Figure 5: Block Model

The block is essentially the collection of transactions which are either mined (PoW) or signed (PoA) by the relevant nodes in the network. The number of transactions included in a single block determine the size of the block, which has no upper limit on the ethereum public blockchain infrastructure.

## 2.4. Smart Contract Structures

Manufacturing related data can be large, diverse, dynamic and proprietary. Hence to allow navigation of the data and selective sharing of data pertaining to an organization, we have designed 3 smart contract representations to represent the relationships between various participants (Figure 6). They are the 1) Global Registrar Contract (GRC); 2) Participant Historical Event Contract (PHEC); and 3) Participant Relationship Contract (PRC).

**Global Registrar Contract (GRC):** This global contract maps each participant – humans, virtual agents and even manufacturing machines to their ethereum address identity. An instance of an ethereum address would be in the form a hexadecimal string such as '0xdc6e86704f6589a4c01cad769581474'. Policies with regards to adding new participants can be coded into the GRC for regulating the addition or modification of existing ones. The contract can also include functions which allow others on the network to pull manufacturer capabilities of a particular entity (Figure 7). The GRC also has code to map to the historical record contract which enables tying together the identity of participants and its historical record.

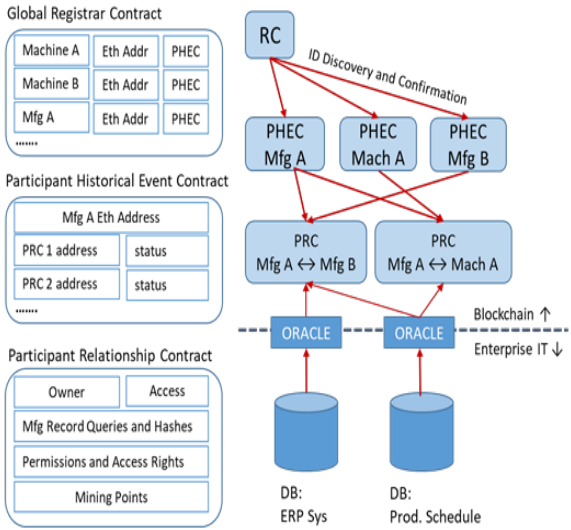

Figure 6: Smart Contract Structure in FabRec. On the right, a subset of the relationship graph between various nodes on the chain

```
pragma solidity ^0.4.16;

contract GRC{
    struct machine{
        //uint index;
        string mname;
        string mac;
        bool status;
        uint availableTime; //time machine is available in minutes
        uint mRate;
        //int[][2][2] signatures;
    }
    struct machOwner{
        //address madd;
        machine[] M;
        bool isExists;
        uint nom;
    }
    mapping(address=>machOwner) public vendor;
    mapping(address=>machine) M_address;
    ...
    ...
}
```

Figure 7: Sample GRC Smart Contract Structure

**Participant Historical Event Contract (PHEC):** This contract belongs to each participant in FabRec, with functions that allow the retrieval of historical events related to the participants with regards to its manufacturing history. This is analogous to the retrieval of credit report of US consumers from the credit agency bureaus. It also contains a record of relationships that a particular participant has with other nodes on the chain. For example, manufacturers who have had relationships with machine vendors will have an established relationship with the transfer of a physical asset, such as a highly specialized equipment (Figure 8). Hence, it would contain all references to relationships that a participant in the chain has been engaged with. This trail of events can be retrieved to help establish reputation and eventually impute trust between potentially new participants from the verified event history.

```
function getMachineInfo(address vadd,uint num) public
                                          constant
                   returns (string ,bool,uint,string);
function getNumberOfMachines(address vadd) constant
                                  returns (uint);
function buyHours(address _seller,string _mac, uint _hours) public
                                          checkSeller(_seller)
                                          returns(string);
```

Figure 8: Some function definitions for the PHEC

**Participant Relationship Contract (PRC):** This contract is setup and initiated when one participant on the chain, for example a job shop service provider enters into a relationship with a client, say a user requesting CNC fabrication services. This relationship can also be extended to data related to a job shop service provider that is managed by another service provider participant. Each entry in the PRC contains binding agreements between the participants and meta-data associated with the relationship. A status also indicates if the relationship is current, voided or successfully completed. Details regarding the relationship can be made available to all participants on the chain, provided both concerned participants have agreed to the terms and conditions. The PRC maintains access pointers to detailed data history regarding a particular relationship within a participant's enterprise IT system. Oracles mediate between an enterprise private IT system and the PRC contract to allow transfer of information across trust boundaries. We can possibly extend features within the PRC to allow a mapping of participants who can be granted access rights.

Smart contracts are written in Solidity, a javascript like language [22]. Special and careful consideration must be used when writing smart contracts because they are inherently immutable. The use of computer code to mediate the relationships is perhaps the single most important feature that powers the autonomous interaction among participants within the chain. An example of a smart contract use case scenario within a peer to peer network of service providers, their machines and consumers is as follows: A machine is attached to or represented by a smart contract on a network. A customer needs a part machined and is in search of a supplier who can meet the customer's specifications of price, quality and time. A contract issued by the customer searches through the GRC, obtains a list of contract addresses that match capability requirements. Further filtering to a selected choice can be undertaken by going through the historical events maintain by the PHEC contracts for a suitable manufacturer. Once a relationship is established through the issuance of a new PRC contract between the customer and the fab service provider, the customer's smart contract can send funds to a smart contract that represents the machine and the machine could decide based on the criteria of the job and proposed funds whether to accept or deny the offer. If the offer is accepted, the machine could queue the job and a receipt of this transaction will be stored on the block.

*2.5. FabRec System Nodal Data Exchange*

We have designed FabRec components to potentially integrate with existing manufacturing ERP and MES systems infrastructure that already exists at various large and medium manufacturer enterprises. We assume that these existing IT systems are trustworthy and they manage their in-house databases that record detailed information with regards to part manufacturing. As shown in Figure 9, our design introduces four software components: 1) A Machine's Virtual Twin Library; 2) Ethereum Client; 3) Nodal Database Sentry through MongoDB and 4) FabRec Blockchain View Manager. Each component has a specific purpose: The virtual twin library enables machine communication with the digital twin/virtual manufacturing machine built on top of a NoSQL database as described in [21]. The Ethereum client is hosted at the client to enable interaction with the blockchain. The Nodal Sentry can be envisioned as a program enforcing cryptographic proofs that the data coming from the VMM has not been tampered with. The blockchain manager can be thought of as an indexing and visualization server to keep track of the transactions on the blockchain in a human readable format (similar to etherscan.io but for machine data.)

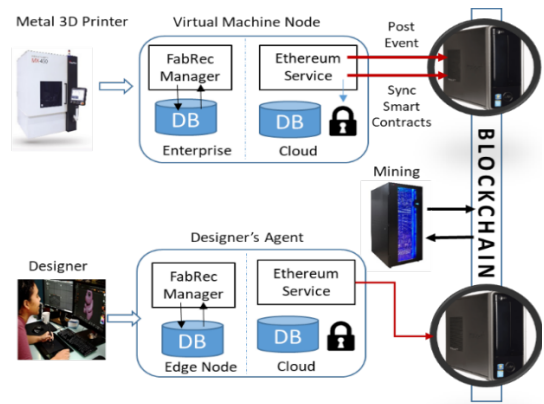

Figure 9: Sample System Data Exchange between sample participants (designers and machines)

We provide a prototype implementation of these various components that integrates with a MongoDB database and managed through a user interface. Also, any database backend and user interface implementation can be utilized as a plug-n-play feature as long as they can interface through API calls to the relevant smart contracts on the blockchain.

It must be noted that FabRec does not contain detailed in-process data related to the fabrication for the part. For instance, metrology data regarding the first article inspection of a part is not stored on FabRec, but rather on the manufacturer's in-house IT system. If permissioned, FabRec will only contain pointers to the data contained within the manufacturer IT system and this exchange of data is made possible through oracles that mediate between the blockchain databases and an enterprise IT system (as shown in Figure 6). An oracle [23] is a piece of code which allows the blockchain to interact with elements outside the blockchain. The oracle may be custom written to interact with the smart contracts existing on the blockchain. Since the smart contract are self-executing, the oracles can provide the data required for computation by the smart contracts.

# 3. Prototype Implementation and Evaluation

## 3.1. Experimental Infrastructure

To simulate a decentralized system, we set up a blockchain network comprising 4 separate computers acting as participants of the blockchain representing various fabrication service providers, couple of nodes containing a copy of the blockchain with two of them assigned as a miner node. We used ethereum as the blockchain platform for its ease of programming smart contracts and the ethereum virtual machine which is written in a Turing complete language. Two computers act as miners on our network which is responsible for validating transactions on network. One of the nodes is connected to an Arduino which interfaces with the blockchain network through an Oracle, which is essentially a python script. This script keeps checking the smart contract via API calls to the JSON-RPC. Transactions can be done through the geth console and through python scripts. Non-computation intensive python scripts allow single board computers to read and write events at low costs. Each of the participant nodes also have an entry written into the Global Registry Contract (GRC) which initiates a Participant Historical Event Contract. A CNC machine (PocketNC, MT, USA) through its MachineKit interface (a modified form of LinuxCNC) running within its Beaglebone Black hardware system was also connected to the network and identified as a participatory node on the decentralized network. Scripts written within the Machinekit OS send critical events (such as Machine ON/OFF/Working status) to its corresponding virtual twin, which then sends event transactions to the network. Mining nodes would verify the authenticity of the event and the added to the current blockchain. Events can be batched up to improve efficiency of the data payload that enters the block. All main nodes on the network achieves parity and consensus to ensure that each node has a copy of the current blockchain. Our setup was intended to demonstrate how a smart contract can check for certain events relayed by the CNC machine (ex, a condition such as continuous 8hr operation) and then the contract triggers a command sent to another physical device, which in our case is a simple LED light attached to an Arduino board. It is important to note that both devices need not be present on the same organization's network.

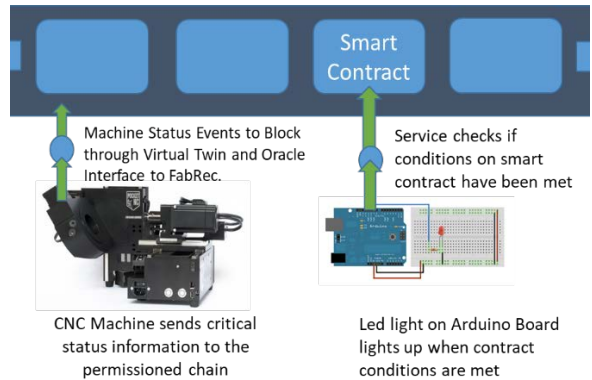

Figure 10: Physical devices interacting on the FabRec chain through the Smart Contract, specifically the Participant Relationship Contract (PRC).

As shown in Figure 10, the smart contract is setup to accept bids from different entities in the form of transactions. The contract was written in a manner to allow transactions with a specific string coded in hexadecimal format. The oracle keeps checking all the transactions with the smart contract and looks for the string that we want. As soon as such a string is identified by the smart contract, the oracle sends a serial signal to the Arduino and the LED lights up. Thus, we were able to demonstrate the use of blockchain for interacting with physical objects. Every node on the network can send bids to the smart contract without knowing the identity of the participant or having to install any special software.

## 3.2. Proof of Authority (PoA) Consensus Algorithm

This is a new consensus algorithm, specifically developed to overcome the limitations of the Proof of Work and Proof of Stake consensus algorithms and is meant only for permissioned blockchains. Here, a number of pre-approved authority nodes, similar to miners on a public chain, approve the set of blocks that must be appended to the current set of blocks in the chain. The authority nodes independently verify the contents of the block, vote on the block and then seal the blocks to the chain network. We have followed the Clique PoA consensus algorithm for our implementation shown in Figure 10 [20].

*3.3. Results*

FabRec provides stakeholders within the permissioned network an immutable log of critical events pertaining to a manufacturer. This information can be made accessible to others on the network depending on access rights set by the manufacturer. The relationship contract between participants, say a manufacturer and their clients or a manufacturer and its machines can be enabled. Below, we consider interoperability and appropriateness of consensus algorithm.

Machine specific adaptors were written to allow the extraction of data from the controllers operating the Pocket CNC machine. MTCONNECT can play a major role in streaming data from the machines to its appropriate virtual twins, which then interface with the blockchain. The LED light on the microcontroller lit up when the specific conditions within the smart contract was met. This shows autonomous decision making and interaction of two physical devices on the blockchain network. The conditions within the smart contract can be configured as long as they are agreed to by the two parties involved. Both PoW and PoA consensus algorithms were tested on our network. Due to limitations of computing power, the difficulty level for the PoW algorithm was set to minimal to ensure completion of the mining of blocks on the network. PoW is inefficient in permissioned blockchains. PoA consensus is more suited to a permissioned chain as it is much faster since participants on a chain are already trusted and their identities are known. PoA through the Clique protocol allows more configuration ability with respect to block size and latency between successive authorizations. Authority nodes can be configured to add as many verification and validation checks as desired by participants on the network. Running the algorithm is also not computationally expensive making it easier to implement.

To test the two consensus algorithms with the machine data generated in our lab, we utilized the public testnets made available on the ethereum blockchain called Ropsten (PoW) and Kovan (PoA) testnets. These testnets come the closest to the performance of the actual Ethereum blockchain. An important point to note here is that "performance" of a contract in terms of execution speed is not relevant in this context. The contracts are executed simultaneously by every miner node in the blockchain.

The only parameters affecting the "performance" of the contract would be the speed of the mining computer's processors. While Solidity, is a Turing complete language, it is not intended to do heavy computation since every single computational process costs a certain amount of 'gas' payable to the miner for carrying out the computations. However, we can measure the performance of the validation protocols such as PoA or PoW. For the purposes of our study, we analysed two metrics: The time it takes for a transaction to be included in a block on the blockchain and secondly, the time it takes to reach a certain number of confirmations in the chain.

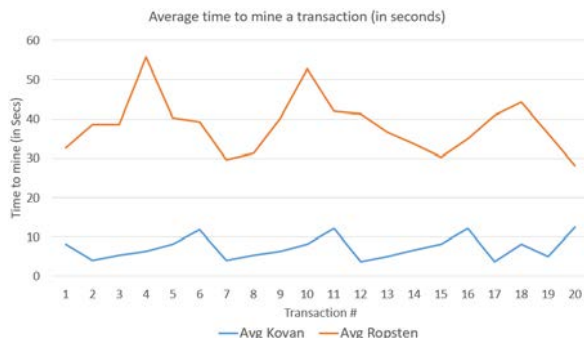

Figure 11: Time to mine a transaction using the PoW and PoA consensus mechanisms

For the first experiment, we wrote a simple representative contract to change the state of the LED machine. For the first test, we looked at comparing the time for inclusion of transactions in a block in the Ropsten and Kovan test blockchain. This would affect the time for the mining of a specific transaction. Till the transaction is mined, it cannot be included in the public chain. For this, we deployed our contract on the Ropsten and Kovan test networks established by the Ethereum Foundation. This was followed up by invoking the contracts randomly 20 times on each testnet. This process was repeated thrice. The resulting times for mining the transactions are shown in Figure 11.

However, just the mining of the block doesn't guarantee the permanence of the transaction in the blockchain. The longer the chain of blocks after a particular transaction is mined, the more immutable and secure the chain becomes. Therefore, for a particular transaction to be un-orphaned (excluded from the main chain due to a potential fork in the future), it is crucial to ensure that a certain number of blocks have been mined. We tested the time it takes for the block to reach 12 confirmations in the methodology noted above. The results are shown in Figure 12.

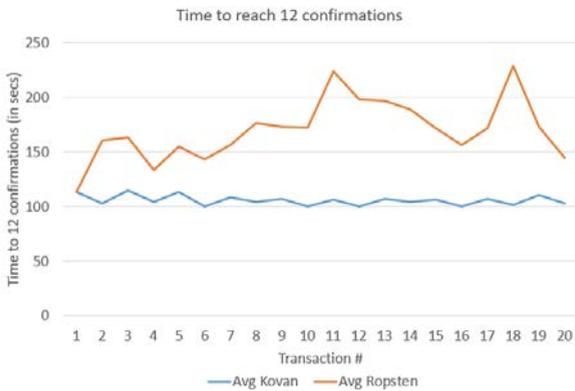

Figure 12: Time to reach 12 confirmations in PoW vs PoA consensus mechanisms.

Table 1: PoA and PoW analysis for time to mine

|  | Avg time | Std Dev |
|---|---|---|
| PoA (Kovan) | 7.25s | 4.91s |
| PoW (Ropsten) | 38.34s | 7.72s |

Table 2: PoA and PoW analysis for time to 12 confirmations

|  | Avg time | Std Dev |
|---|---|---|
| PoA (Kovan) | 105.57 | 17.31 |
| PoW (Ropsten) | 170.01 | 45.79 |

In both cases, the Proof of Authority works better in terms of time for inclusion of transaction in a block as well as the time to confirm a block efficiently. The average time taken to mine one transaction is nearly 7.25s in a PoA mechanism vs ~38s in a PoW mechanism for our tests (Table 1). The standard deviations showed by the PoW is also much higher compared to the time taken to confirm transactions in PoA (Table 1 and 2).

However, this does not mean that the PoW consensus is worse than PoA. This shows that the consensus mechanism has to be chosen depending on the application and the environment in which the blockchain needs to be implemented. For example, if a blockchain system for provenance tracking across a large supply chain is being implemented, a PoW mechanism assures us of higher security since all participants need not trust each other completely, unlike PoA where participants need to trust a few nodes to sign their transactions in blocks. On the other hand, if a blockchain system is implemented for verification of machine states such as states of the CNC machine are to be implemented within a consortium where participants trust one another, the PoA will be sufficient. Also, we can note that in both cases, the time to confirm transactions (as seen in the Fig 12) plays a crucial role. Business decisions should be taken after a certain number of confirmations (say 7 to 12) have taken place since the probability of the block containing transactions of interest getting 'uncled' [10] is very low. This means that the blockchain is not particularly suitable for real time data transfer and decision making. In either case, we have shown that the blockchain system assures provenance of data and transparency at low costs by sacrificing the performance of the network deliberately.

## 4. Industrial Implementation Challenges and Future Work

Interest in blockchain has seen tremendous amounts of investment, particularly with the rise in the value of cryptocurrency such as Bitcoin and Ether [25]. In manufacturing, there is no notion of currency being traded, but rather the flow of physical assets and digital meta-data surrounding the parts making its way through various suppliers. There are several technical, business and legal challenges to its wider scale adoption in manufacturing. One, the very nature of blockchain necessitates that this shareable database cannot be owned by any single central authority. Therefore, the natural question that arises is, how are the nodes on the network managed and can it reliably be trusted against nefarious attacks by external participants? New business models must be designed to incentivize the participation of various manufacturer stakeholders on the network, ensuring that there is no single group of entities that control the network.

Smart Contract structures despite being proposed in early 1990s, still lack several critical features to allow the enforcement of contract terms and conditions, specifically for manufacturing. New advances in the form of 'cryplets', which separate logic from the underlying data layer allows more robust and stable execution [24]. Implementing business logic in the form of computer code for complex manufacturing environments may not always be feasible, requiring both off chain and on-chain execution. This will require close collaboration with legal entities to ensure viable implementation. Sharing of manufacturing data can be a sensitive topic for many manufacturers. Lack of network connectivity of the machines both due to security concerns and lack of infrastructure prevent the wider scale adoption of blockchain participants to the very edge of the manufacturing network. Changing current practices of information and data sharing by participants can be a significant shift away from the traditional mindset. However, it does present new opportunities for manufacturing organization to participate in the digital economy.

We have identified several thrusts for future work that must be carried out to expand the system. First, as noted in the industrial implementation challenges, a coordinated effort of engagement must be initiated with manufacturing stakeholders across the industry – designers, job shop service providers, machine builders, regulators, certifiers, public policy and corporate law. Second, significant semantic and ontology based information modeling frameworks must be built to enable smart manufacturing marketplace of buyers, agents and service providers. Third, computing reputation and evaluating trust based on manufacturing events recorded on the chain is still emerging and will be needed for the automatic selection and negotiation of job contracts. Finally, since data is now available on the chain, artificial intelligence can now be encoded into agents working on behalf of the participants which if designed well, can lead to efficient and fair marketplaces.

## 5. Conclusion

The FabRec prototype provides a proof of concept system linking computing nodes and physical devices such as the Arduino/Raspberry PI and a basic CNC machine to demonstrate the feasibility of connecting these nodes on a decentralized and interoperable network. We have used the Ethereum smart contracts to autonomously initiate commands being given to the Arduino system based on events recorded by the physical machine on another part of the network. We demonstrate data models for the various smart contracts that must be built within FabRec to enable smart manufacturing marketplaces of designers and machines. There is much future work to be done to combine database principles, cryptography, network topology and microeconomics into the digital manufacturing application domain.

## Acknowledgements

We thank the financial support from NSF Cybermanufacturing #1547105 (AA, BC and BS) and the Edward P. Fitts Fellowship (MH).